# Open Data Platform for Knowledge Access in Plant Health Domain : VESPA Mining

Nicolas Turenne[1], Mathieu Andro[2], Roselyne Corbière[3], Tien T. Phan[4]

1 Université Paris-Est, LISIS, INRA, F-77454 Marne-La-Vallée, France
2 INRA, DIST, F-78026 Versailles, France.
3 INRA, UMR1349 IGEPP, F-35653 Le Rheu, France
4 Vietnam National University of Agriculture, Hanoi, Vietnam

Introduction

Important data are locked in ancient literature. It would be uneconomic to produce these data again and today or to extract them without the help of text mining technologies. Vespa is a text mining project whose aim is to extract data on pest and crops interactions, to model and predict attacks on crops, and to reduce the use of pesticides.

A few attempts proposed an agricultural information access. We can find systems with socio-semantic approach (Turenne & Barbier, 2004), or systems with thesaurus-oriented approach like (Milne et al, 2006; Bartol, 2009; Vakkari, 2010; Laporte et al, 2012). But these systems are not driven by concrete usages like the Vespa Mining platform.

Another originality of our work is to parse documents with a dependency of the document architecture (sections: chapters, subchapters...). Lots of works has been done to retrieve named entities like genes, persons, organization... in texts with a good efficiency (Riloff, 1996; Roth & Yih, 2002; Carpenter, 2007; Surdeanu et al, 2011; Krieger et al, 2014). Some state of the art tools are even available (http_Lingpipe, 2015; http_SNER, 2015) but in these approaches, entities are searched in a sentence and relations are often searched in a same sentence. Our approach is different, we search relations in a same section. Our approach (Turenne & Phan, 2015) is based on dictionary matching and on rules defined by users using a tool called Unitex (htt_Unitex, 2015) to detect entities and document design together with cooccurrence analysis to detect relation. High efficiency relation extraction reduces time of browsing in documents to identify relevant information.

1- A data source, a printed bulletin: "Avertissements Agricoles"

Since the 60s, the French Regional Plant Protection Services (SRPV) and Groupings Protection against Harmful Organisms (GDON) have published printed bulletins organized by French regions and by crops. These bulletins are about the arrival of pests, fungi, diseases and they advice about how to fight them. Avertissements agricoles, used by farmers and their consultants, contain many data about species, pesticides, meteorologic, geographic and also specific observational data. These data were collected by a network of agents and experts in charge to control the evolution of diseases, insects and other pests of crops. Avertissements agricoles are a source of historical data that will provide a better understand of the effects of climate change on pests and diseases and agricultural practices. The most complete collection of this publication was found at the National Library of France (BnF) where, like every other journals, it was preserved under legal deposit. With funding from French National Institute For Agricultural Research (Inra), SMaCH pour Sustainable Management of Crop Health metaprogram, nearly 50 000 pages of this bulletin were digitized by the National Library in 2013-2014. Then, OCR and controlled segmentation was obtained from an external provider. With funding from the French Ministry of Research (BSN call for digitization projects), gross OCR is now being corrected by an external provider to obtain corrected OCR (98 %) and light encoding of texts structures (Text Encoding Initiative). A good enough quality of OCR will improve text mining extractions.

We also plan to use crowdsourcing plateforms to obtain human annotations. This would allow us to enrich the project annotation resources and dictionnaries. So, this would improve the efficiency of text mining extractions. This would also allow us to evaluate the quality of text mining annotations compared to human labor, for precision and recall. We will use a sample of 200 bulletins and wil confront annotations from, at least, two crowdsourcers to better control human annotation quality.

2- Information search web-platform

2.1 Information extraction

A standard BSV document consists of warning section about crops and its bioagressor. From mining point of view a simple word indexing is not sufficient because not enough accurate in term of end-user objective, and term extraction is not accurate because BSV is written in natural language including lots of common noun phrases. Named entity extraction and relation extraction fits better user goals, and the BSV type of corpus is well suitable for information extraction due to the natural science content linked to diseases, species, locations and time. Due to a specific document structure in section we used three assumption to help our information extraction task :

1- Relevant items and their relationships can occur to different architectural part of a document
2- Our second assumption is that relationship between named entities are explicit and able to be catched by shallow parsing of tokens in the text.
3- Items found in texts can be interpreted by an end-user and ts its usage needs

We adopt a pattern-matching approach with local grammar rules and a domain thesaurus. The 11 concepts concern the type of extraction that are the following: plant developmental stages, publication times, issue numbers, regions, damage, crops, pests, diseases, auxiliary insects, chemicals, and climat. From these concepts only two kind of relationship interest our users: crop-pest and crop-disease. Other concept can highlight relationships. Some concept concern all cited relationships in a same document such as publication time, issue number or region. The other are more specific to an instance of a relationship. But we focused mainly about intances pest-crop-damage and pest-disease-damage. This is an exemple :

<div align="center">wheat-rust-12% of infested parcels</div>

We used a classical unsupervised learning approach called cooccurrence analysis but in synergy with document structure analysis.

As document structure analysis plays an important role we used heuristics involved in the relation extraction:

1- Main entity : A target entity occur in a specic title or subtitle (beginning of a paragraph or a section).
2- Header: Different entities occurs in the header of the document (rst lines).
3- Avoid section: Some paragraphs begining by a specic title can contain entity but not associated to a main entity or contextual information

Relation extraction takes as input itemsets to identify relations the export from Algorithm 1, and plays with the three heuristics. Heuristics 1 set that some entities are main entities (i.e. a category is a chosen as a target) and we seek relations for these entities. Heuristics 2 sets that target entities are declared in header sections (titles, subtitles) and heuristics 2 declares that some sections are non-relevant, we called them avoid sections and they can be specied by a begining phrase and can end by the end of document of another phrase. Algorithm 2 describes how relation extraction is implemented. x.ent implements also a class of algorithm to detect relation without heuristics in case, a document only consists of paragraphs (i.e a tweet, an email or a news).

## 2.2 Web portal

A network of users in agronomy and plant epidemiology fields is concerned by the information system we aim to implement as a service. Four users, specialists about potatoes and wheat helped us to control user-interface building and search options with a minimal investment into the system handling for a new user.

A global keyword query is available in the graphical interface to bootstrap a query according a crop, a disease or a pest only , or a combination crop-disease or crop-pest. Clicking on the query execution button, two results, with the matching documents, are exported, below the keywords query (Figure 1). The first is the flat list of documents sortable by date, the second is the map of French regions on which hits of documents are visible for each region.

Fig. 1 User interface of the web-platform.

The user can refine the results by two means. Playing with the main query with a textbox specifying a free word acting a a filter if present in any part of result document, and with publication date range. Other options are available on the map, if a simple species is specied in the query , by clicking on the mouse button list of other concept linked to the query species can be selected (for instance if a crop is in the query , on the map for a given region the list of pest and disease linked to the crop are availble for refinement). If a relationship is in the query then the mouse button gives information for each document about damages for a given region. We can mention four cases of usage with main query and renement:

    A-  main query is relation crop-disease, renement damage and region on map.
    example: Crop=wheat, Disease=rust, On map : risk assessment, region=Burgondy

(in French: Plante=blé, Maladie=rouille, on map : region=Bourgogne)

B- main query is crop, renement pest (relation crop-pest) and region on map.
example: Crop=rapeseed, On map : Pest=mouche du chou, region=Centre,
and document sorting by date
(in French: Plante=colza, Maladie=rouille, on map : region= Centre)

C- main query is disease, renement pest (relation crop-disease) and region on map.
example: disease=potato late blight, On map : crop=potato, region=Burgundy
(in French: Plante=pomme de terre, Maladie=mildiou, on map : region=Bourgogne)

D- main query is pest, renement pest (relation crop-pest) and region on map.
example: pest=fly, On map : crop=wheat, region=Midi-Pyrenees
(in French: Plante=blé, Ravageur=mouche, on map : region= Midi-Pyrénées)

At present 4 users specialists about potato and wheat take benefit from the database and platform-as-service. More epidemiologists and agronoms, from the integrated crop protection network (in French: reseau PIC - protection integree des cultures) including 400 subscribers, are potentially interested in using this web platform. Risk analysis in a sociological point of view is also possible.

## 2.3 Availability of computing tools

Information extraction tool (x.ent) is written in perl and R languages and available from CRAN website for download : http://cran.r-project.org/web/packages/x.ent/index.html. Configuration is done by a browser displaying a file with different field to fullfil as corpus path, local grammar paths. The x.ent tool can be applied to any kind of corpus independantly from the web portal. The Vespa Mining web portal is at moment freely available from this temporary web site : http://vespa.cortext.net.